\documentstyle[prb,aps,amssymb,multicol,epsf,psfig,pstricks]{revtex}

\newcommand{\onecolm}{
  \end{multicols}
  \vspace{-3.5ex}
  \noindent\rule{0.5\textwidth}{0.1ex}\rule{0.1ex}{2ex}\hfill
}
\newcommand{\twocolm}{
  \hfill\raisebox{-1.9ex}{\rule{0.1ex}{2ex}}\rule{0.5\textwidth}{0.1ex}
  \vspace{-4ex}
  \begin{multicols}{2}
}

\begin{document}
\def\CC{{\rm\kern.24em \vrule width.04em height1.46ex depth-.07ex
\kern-.30em C}}
\def\P{{\rm I\kern-.25em P}}
\def\RR{{\rm
         \vrule width.04em height1.58ex depth-.0ex
         \kern-.04em R}}
\def\id{{\rm 1\kern-.22em l}}
\def\beq{\begin{equation}}
\def\eeq{\end{equation}}
\newenvironment{eqblock}[2]{\beq\label{#2}\begin{array}{#1}}{\end{array}                                \eeq}
\newenvironment{neqblock}[1]{\[\begin{array}{#1}}{\end{array}\]}
\newcommand{\beqb}{\begin{eqblock}}
\newcommand{\eeqb}{\end{eqblock}} 
\newcommand{\nbeqb}{\begin{neqblock}}
\newcommand{\neeqb}{\end{neqblock}} 
\newcommand{\bigfrac}[2]{\mbox {${\displaystyle \frac{ #1 }{ #2 }}$}}
\newcommand{\eps}{\varepsilon}
\newcommand{\sDelta}{{\scriptstyle\Delta}}
\def\beqa{\begin{eqnarray}}
\def\eeqa{\end{eqnarray}}
\renewcommand{\i}{{\rm i}}

\title{Fermionic long range correlations realized by particles obeying deformed statistics}
        
\author{Andreas Osterloh$^{1,2}$, Luigi Amico$^{2}$, and 
Ulrich Eckern$^{1}$}
\address{$^{1}$Institut f\"ur Physik,Universit\"at Augsburg, D-86135 Augsburg, Germany}
\address{$^{2}$ Dipartimento di Metodologie Fisiche e Chimiche (DMFCI), Universit\'a di Catania, viale A. Doria 6, I-95129 Catania, Italy.\\
Istituto Nazionale per la Fisica della Materia, Unit\'a di Catania, Italy.}

\maketitle
\begin{abstract}
Deformed exchange statistics is realized in terms of electronic 
operators. This is employed to rewrite Hubbard type lattice models
for particles obeying deformed statistics (we refer to them as  deformed models) 
as lattice models for electrons.
The  resulting models show up gauge--like modulations in 
the hopping processes, which induce long-range correlations in the lattice.
The conditions  for the Bethe ansatz solvability of the latter 
are interpreted as   restrictions imposed on   
the statistics to be compatible with the Bethe ansatz solvability of the deformed models.
It is found that solvable deformed models
are not unitarily equivalent to fermionic models if  
the exchange of particles with the same spin-orientations is
deformed. 
Statistics deformations, where the exchange relation of two particles is
influenced by the presence of other particles, cannot be realized 
by  fermionic operators.

\end{abstract}
\begin{multicols}{2}

The statistics of degrees of freedom drastically affects the physical
properties of a many-particle system.
Besides bosonic and fermionic statistics, 
a continuous family of intermediate statistics serves to explain 
important effects involved in two or one dimensional physics.
Remarkably, in $D=2$, excitations in the fractional quantum Hall 
effect can be described as anyons~\cite{HALL,MYRHEIM}. 
One-dimensional systems can occur either because only one-dimensional
dynamics is allowed
in the system (even if the system lives in higher dimensions) or 
because the samples are indeed one-dimensional
(like quantum wires, carbon nanotubes, systems with charge density wave order, etc).
In the present paper we shall focus on deformed statistics in one dimension.
\\
Defining arbitrary statistics in one dimension exhibits several 
peculiarities~\cite{POLYCHR}. In particular, 
imposing the statistics in one dimension can be interpreted as 
a ``continuity condition'' on the wave function (arising from the set 
of coordinates such that two or more particles coincide),
fixing  its symmetry.
One dimensional fractional statistics  arises since this  constraint on the wave function 
can be imposed arbitrarily.
Explicit realizations of $1D$ fractional statistics quasi particles, 
formulated in ``first quantization'',
are the eigenstates of Calogero--Sutherland models.  

In  Refs.~\onlinecite{AMOSECK} and \onlinecite{OSAMECK}, the notion of Deformed
Exchange Statistics (DES) was defined as a specific deformation of
electronic commutation rules (in second quantization).  
Mathematical aspects of this  type of statistics have been also studied 
in Refs.~\onlinecite{Qij-CCR,MARCINEK}.
We applied DES to 
investigate how robust is the solvability by Coordinate Bethe Ansatz 
(CBA)~\cite{BETHE,SUTHERLAND-NOTES} 
of the XXZ and the Hubbard model  with respect to such a modification
of the particle content (we have called those DES preserving CBA
solvability {\em Solvable DES}).
\\
Following  different physics,  
H. Schulz and B. S. Shastry~\cite{SCHULZ} have included long 
range correlations in the Hubbard model 
through a gauge-like modification of the kinetic term of the Hamiltonian.
In Ref.~\onlinecite{OSAMECK} we have extended the 
CBA solution of Schulz-Shastry models 
to models where more general forms of the correlation have been 
considered; characterizations focusing on
CBA solvability of such kind of correlated models have been reached.
The Bethe  equations arising from solvable Schulz--Shastry type models show that 
the long range correlations induce a twist in the boundary 
conditions.

In the present paper we show that the solvable 
deformed   Hubbard model 
(without Schulz--Shastry type correlations) 
is equivalent to adding correlations similar
to those discussed by Schulz and Shastry\cite{SCHULZ}
to the undeformed Hubbard model. 
We prove this  realizing DES operators 
by composites of electronic operators.
Using the results of
Ref.~\onlinecite{OSAMECK-GEN}  we can characterize DES which preserve 
the  CBA solvability of the undeformed  model (see Eq~(\ref{INT-DES}).

Particles obeying DES have creation and annihilation operators obeying 
the following deformed commutation rules: 
\begin{eqnarray}\label{ANY1}
f_{j,\sigma}^\dagger f^{}_{{ k},\sigma'} + {\cal Q}_{j,k}^{\sigma,\sigma'}\,  
f^{}_{{ k},\sigma'} f_{{ j},\sigma}^\dagger &=& \delta_{j,k}\
\delta_{\sigma \, \sigma '} ,\\
\label{ANY2}
f^{}_{{ j},\sigma} f^{}_{{ k},\sigma'} + {\cal Q}_{k,j}^{\sigma',\sigma} \,  
f^{}_{{ k},\sigma'} f^{}_{{ j},\sigma} &=& 0 ,
\end{eqnarray}
Such an algebra is consistent for non trivial ${\cal Q}_{j,k}$ 
if:
\begin{equation}
{\cal Q}_{j,k}^{\sigma\sigma'} = ({\cal Q}_{k,j}^{\sigma',\sigma})^{-1}  = 
({\cal Q}_{k,j}^{\sigma', \sigma})^\dagger \; .
\label{desCONS1}
\end{equation}
Furthermore it will be postulated
\begin{equation}
[ f^\dagger_{j,\sigma}\, ,\, {\cal Q}_{j,k}^{\sigma,\sigma'}] \ 
=\ [ f^{}_{j,\sigma} \, ,\, {\cal Q}_{j,k}^{\sigma,\sigma'} ] \ = \ 0 \; .
\label{despostulate-rel} 
\end{equation}
The operators $\nu_{{ j},\sigma}\doteq 
f_{{ j},\sigma}^\dagger f_{{ j},\sigma}$ are the particle-number 
operators. 
Note that the relation (\ref{ANY1}) is formally analog to quon commutation
rules (called $q$--CCR in Ref.~\onlinecite{Qij-CCR}), where relation (\ref{ANY2}) however cannot hold except for the
fermionic and bosonic case~\cite{WU}. 
Here, in contrast to the quon algebra, the deformation parameter
depends on site and  spin indices $(j,\sigma| k, \sigma')$ 
(this  statistics is of the type  called $q_{ij}$--CCR 
in Ref.~\onlinecite{Qij-CCR}). \\   
The Equations (\ref{desCONS1}) and (\ref{despostulate-rel}) guarantee
that the particles are representations of the permutation group $S_N$
and ensure the validity of the standard commutation relations:
\(
[ \nu_{i,\sigma},  \nu_{j,\sigma'} ]=0
\), 
\(
[ \nu_{i,\sigma}, f_{j,\sigma'}^\dagger ]=
\delta_{i,j}\delta_{\sigma,\sigma' }f_{j,\sigma'}^\dagger\), and  
\(
[ \nu_{{i},\sigma}, f_{{ j},\sigma'}]=-
\delta_{i,j}\delta_{\sigma,\sigma' }f_{{ j},\sigma'}\).
This provides a well defined Fock representation of the algebra
defined in Eqs.~(\ref{ANY1}) and (\ref{ANY2}). 

The key observation is that the DES defined by
Eqs.~(\ref{ANY1})--(\ref{despostulate-rel}) are representable using
operators which are composites of electronic creation/annihilation
operators. 
\beq
f^\dagger_{i,\sigma}:= c^\dagger_{i,\sigma}\exp i\bigl[
\Delta^{\sigma;\lambda}_{i;l} n_{l,\lambda} + 
\Delta^{\sigma;\lambda,\mu}_{i;l,m} n_{l,\lambda} n_{m,\mu} + \dots
\bigr] , 
\label{realization}
\eeq   
where, without loss of generality, $\Delta$ vanishes for any two
coinciding index pairs and is symmetric in exchanging arbitrary
index pairs behind the semicolon.
The number operators remain unchanged after this realization: 
$\nu$ transport into the number operators $n$ for fermions.
The resulting deformation parameters are
\beqa
{\cal Q}_{j,k}^{\sigma, \sigma'} &=& \exp i\bigl[ 
(\Delta_{k;j}^{\sigma';\sigma} - \Delta^{\sigma;\sigma'}_{j;k} ) + \nonumber \\
&&\ + 2(\Delta^{\sigma';\sigma,\mu}_{k;j,m}-\Delta^{\sigma;\sigma',\mu}_{j;k,m})n_{m,\mu} +
\dots
\bigr]\label{QDEFORM}.
\eeqa
From Eq.(\ref{QDEFORM}) it is seen that no deformation occurs {\it
iff} $\Delta$ is totally symmetric in the index pairs. Thus, it will
be assumed being antisymmetric in exchanging one index-pair behind the
semicolon with the index pair in front of it. This already implies
that $\Delta$ vanishes if it has more than two index pairs. 
The deformation parameter expressed in terms of the parameters $\Delta$'s in 
Eq.~(\ref{realization})
is
\beq\label{Q-Delta}
{\cal Q}_{j,k}^{\sigma, \sigma'} = \exp 2{\rm
  i}\Delta_{k;j}^{\sigma';\sigma} = \exp -2{\rm
  i}\Delta_{j;k}^{\sigma;\sigma'}.
\eeq
As a consequence it is impossible to represent correlated 
DES (that is DES with  ${\cal Q}^{\sigma,\sigma'}_{j,k}$ being a
functional of $n_{m,\mu}$) through the realization~(\ref{realization}). 

Before we continue the discussion of DES, 
it is worthwhile noting that a totally symmetric part in $\Delta$
creates correlated hopping (without changing the statistics of
the particles). For this reason, the fermionic hopping term, which is
created by the realization~(\ref{realization})  of the degrees of freedom of the 
deformed Hubbard model is calculated for general $\Delta$:
\begin{equation}\label{ourHAMILTON}
H=  - {t}\, \sum_{i,\sigma } 
\, \left ( f_{{ i},\sigma}^\dagger f_{{i+1},\sigma} + 
{\rm h.c.} \right ) +
U\, \sum_{ i}\nu_{{i},\uparrow} \nu_{{ i},
\downarrow} \; , 
\label{MONRAMODEL}
\end{equation}
where $f^{}_{i,\sigma}$, $f_{i,\sigma}^{\dagger}$ 
($\sigma \in \{\uparrow,\downarrow \}$, or equivalently $\sigma\in
\{1/2,-1/2 \}$), obey the  deformed relations~(\ref{ANY1}) and
(\ref{ANY2}). The two contributions in the  
Hamiltonian are the hopping term (the $t$--term) and the Coulomb 
interaction term (the $U$--term). 
Now we realize the operators $f^\dagger_{i,\sigma}$ through electronic
operators, using Eq.~(\ref{realization}).
Then the $t$--term in~(\ref{MONRAMODEL}) is rewritten as
\beqb{l}{DES-Hop}
f^\dagger_{j+1,\sigma} f^{}_{j,\sigma} =
c^\dagger_{j+1,\sigma} c^{}_{j,\sigma}\exp i \bigl[
-\Delta^{\sigma;\sigma}_{j+1;j} + \\ 
\\
\hspace{5mm} + (\Delta^{\sigma;\mu}_{j+1;m}-\Delta^{\sigma;\mu}_{j;m}
-2\Delta^{\sigma;\sigma,\mu}_{j+1;j,m}) n_{m,\mu} + \\
\\
\hspace{5mm} +
(\Delta^{\sigma;\lambda,\mu}_{j+1;l,m}-\Delta^{\sigma;\lambda,\mu}_{j;l,m}
- 3\Delta^{\sigma;\sigma,\mu,\lambda}_{j+1;j,m,l}) n_{l,\lambda}
n_{m,\mu} +\\ 
\\
\hspace{2cm} + \dots\bigr].
\eeqb
Taking explicitly account for the subrelevant parts, i.e. the terms
in which the number operator $n_{j,\sigma}$ appears, this is
equivalent to 
\beqb{l}{Symm-Delta-Hop}
f^\dagger_{j+1,\sigma} f^{}_{j,\sigma} =
c^\dagger_{j+1,\sigma} c^{}_{j,\sigma}\exp i \bigl[ \\
\\
\hspace{5mm}
(\tilde{\Delta}^{\sigma;\mu}_{j+1;m}-\tilde{\Delta}^{\sigma;\mu}_{j;m})
n_{m,\mu} + \\ 
\\
\hspace{5mm} +
(\tilde{\Delta}^{\sigma;\lambda,\mu}_{j+1;l,m}-\tilde{\Delta}^{\sigma;\lambda,\mu}_{j;l,m})
n_{l,\lambda} n_{m,\mu} +\\ 
\\
\hspace{2cm} + \dots\bigr],
\eeqb
where $\tilde{\Delta}$ is the same as $\Delta$ except that
$\tilde{\Delta}^{\sigma;\sigma,\dots}_{j+1,j,\dots}=0$ now\cite{OSAMECK-GEN}. 
We find that the solvability conditions obtained in
Ref.~\onlinecite{OSAMECK-GEN} are all fulfilled for $\tilde{\Delta}$
and hence the boundary phases are then given by
\beq
\sum_{j=1}^L \left ( \tilde{\Delta}^{\sigma,\dots}_{j+1,\dots} -
 \tilde{\Delta}^{\sigma,\dots}_{j,\dots} \right ) = 0.
\eeq
Hence we obtain as result that a totally symmetric part can always be
gauged away without residual boundary phase.

Now we continue the study of DES. That means we now restrict ourselves
to antisymmetric $\Delta^{\mu;\nu}_{m,n}$. All parameters with more
than two index pairs are zero.\\ 
Since the electron number operators coincide with the $\nu$ operators,
the $U$--term in~(\ref{DES-Hop}) coincides with the Hubbard Coulomb 
interaction: $U\, \sum_{ i}n_{{i},\uparrow} n_{{ i},
\downarrow}$.
Writing the $t$--term in the form used in the
Refs.~\onlinecite{SCHULZ} and \onlinecite{OSAMECK-GEN}, namely
\begin{eqnarray}
\, \sum_{j,\sigma } 
\, &&\biggl\{ c_{j+1,\sigma}^\dagger c_{j,\sigma} \exp({\rm i}\gamma_j(\sigma))\times \nonumber \\
\ && \times\exp\Bigl[{\rm i}\sum_{l}^{}\bigl(\alpha_{j,l}(\sigma)n_{l,-\sigma}
   +  A_{j,l}(\sigma)n_{l,\sigma}\bigr)\Bigr] + {\rm h.c.} \biggr\},  \nonumber
\end{eqnarray}
we can compare with Eq.~(\ref{DES-Hop}), and the parameters can be
identified being 
\beqa
\gamma_j(\sigma) &=& -\Delta^{\sigma;\sigma}_{j+1;j}, \\
\alpha_{j,m}(\sigma)&=&\Delta^{\sigma;-\sigma}_{j+1;m}-\Delta^{\sigma;-\sigma}_{j;m}, \\
A_{j,m}(\sigma)&=&\Delta^{\sigma;\sigma}_{j+1;m}-\Delta^{\sigma;\sigma}_{j;m}.
\eeqa
The deformed Hubbard model is CBA solvable
if the following conditions are fulfilled~\cite{OSAMECK-GEN}:
\beqa
\label{closedness1}
\alpha_{m,j+1}(-\sigma) - \alpha_{m,j}(-\sigma) &=&
\alpha_{j,m+1}(\sigma) - \alpha_{j,m}(\sigma) , \\ && \nonumber \\
\label{closedness2}
A_{m,j+1}(\sigma)-A_{m,j}(\sigma)&=&A_{j,m+1}(\sigma)-A_{j,m}(\sigma) \\
&& \mbox{for } m\neq j,j\pm 1 \; . \nonumber
\end{eqnarray} 
These conditions for CBA solvability can be expressed directly in terms 
of the deformation parameters ${\cal Q}_{j,k}^{\sigma, \sigma'}$ as
\beq
\bigfrac{{\cal Q}_{j,k+1}^{\sigma, \sigma'}{\cal Q}_{j+1,k}^{\sigma, \sigma'}}
        {{\cal Q}_{j,k}^{\sigma, \sigma'}{\cal Q}_{j+1,k+1}^{\sigma, \sigma'}}= 
1 \label{INT-DES} 
\eeq
for $k\neq j,j\pm 1 \ \vee \sigma\neq\sigma'$.
\\
If the conditions (\ref{INT-DES}) are fulfilled, the Bethe equations for
periodic boundary conditions are
\begin{eqnarray}\label{desBETHEEQ}
&&{\rm e}^{{\rm i} p_j L} = {\rm e}^{-{\rm i}\Phi_\uparrow}
                \prod_{a=1}^{N_\downarrow}
                \frac{{\rm i}(\sin p_j - \zeta_a) -\frac{U}{4t}}
                        {{\rm i}(\sin p_j - \zeta_a) +\frac{U}{4t}}, \\  
&&\prod_{b=1 \atop b\neq a}^{N_\downarrow}
\frac{i(\zeta_a - \zeta_b) +\frac{U}{2t}}{i(\zeta_a - \zeta_b) -\frac{U}{2t}}
=  {\rm e}^{-{\rm i}(\Phi_\uparrow-\Phi_\downarrow)} 
          \prod_{l=1}^N
                \frac{{\rm i}(\sin p_l - \zeta_a) -\frac{U}{4t}}
                        {{\rm i}(\sin p_l - \zeta_a) +\frac{U}{4t}}. \nonumber   
\end{eqnarray}
where the boundary twists are given by
\beq
\Phi_\sigma:=\phi(\sigma)+ \phi^{(1)}_{\uparrow\downarrow}(\sigma) N_{-\sigma} +
\phi^{(1)}_{\uparrow\uparrow}(\sigma)(N_\sigma-1) .
\eeq
They can be written in terms of the parameters entering the
statistics in the following way:
\beqa
\phi^{(1)}_{\uparrow\downarrow}(\sigma)&=&\sum_{j=1}^L 
\alpha_{j,m}(\sigma)= \nonumber \\
&=&\sum_{j=1}^L  
(\Delta^{\sigma;-\sigma}_{j+1;m}-\Delta^{\sigma;-\sigma}_{j;m})=0, \\
&& \nonumber \\
\phi^{(1)}_{\uparrow\uparrow}(\sigma)&=&\!\!\!\sum_{{j=1\atop j\neq m-1,m}}^L 
\!\!\!A_{j,m}(\sigma)+\nonumber \\
&&\quad + A_{m,m-1}(\sigma)+A_{m-1,m+1}(\sigma)\nonumber \\
&=&\sum_{j=1}^L  
(\Delta^{\sigma;\sigma}_{j+1;m}-\Delta^{\sigma;\sigma}_{j;m})+\nonumber \\
&&+\Delta^{\sigma;\sigma}_{m+1;m-1}-\Delta^{\sigma;\sigma}_{m;m-1}
+\Delta^{\sigma;\sigma}_{m;m+1}-\nonumber \\
&&\Delta^{\sigma;\sigma}_{m-1;m+1}+\Delta^{\sigma;\sigma}_{m-1;m}-
\Delta^{\sigma;\sigma}_{m+1;m} \nonumber \\ && \nonumber \\
&=& 2(\Delta^{\sigma;\sigma}_{m+1;m-1}+\Delta^{\sigma;\sigma}_{m;m+1}+
\Delta^{\sigma;\sigma}_{m-1;m}), \\
\phi(\sigma) &=& \sum_{j=1}^L \left ( \gamma_{j}(\sigma) + A_{j,j}(\sigma)
\right )  = \nonumber \\
&=&  \sum_{j=1}^L \bigl[-\Delta^{\sigma;\sigma}_{j+1;j} +
\Delta^{\sigma;\sigma}_{j+1;j}\bigr] = 0.
\eeqa
Thus the total boundary phase is obtained as
\beq
\Phi_\sigma=2(\Delta^{\sigma;\sigma}_{m+1;m-1}+\Delta^{\sigma;\sigma}_{m;m+1}+
\Delta^{\sigma;\sigma}_{m-1;m})(N_\sigma-1).
\label{b-phases}
\eeq
Using Eq. (\ref{Q-Delta}) we get
\beq\label{Q-phases}
\exp \i \Phi_\sigma = {\cal Q}^{\sigma,\sigma}_{m-1,m+1} {\cal
  Q}^{\sigma,\sigma}_{m+1,m} {\cal Q}^{\sigma,\sigma}_{m,m-1}.
\eeq 
We point out that the only non-vanishing phases arise from statistics
deformation for particles having the same spin orientation.

A consequence of this result is that every translational invariant 
uncorrelated DES does not affect the spectrum. This includes purely
spin dependent DES. This can be seen noting that for translational
invariant deformation parameter, i.e. 
${\cal Q}^{\sigma,\sigma'}_{j,k}=:{\cal Q}^{\sigma,\sigma'}_{j-k}$,
the solvability condition (\ref{INT-DES}) must hold without
exception\cite{OSAMECK-GEN}. Thus we have 
\beq
\bigfrac{{\cal Q}_{j-k-1}^{\sigma, \sigma'}{\cal Q}_{j+1-k}^{\sigma, \sigma'}}
        {{\cal Q}_{j-k}^{\sigma, \sigma'}{\cal Q}_{j-k}^{\sigma, \sigma'}}\stackrel{!}{=} 
1 \label{INT-DES-TRANS} 
\eeq
for arbitrary $j$, $k$, $\sigma$ and $\sigma'$.
For $\sigma=\sigma'$ and $j=k+1$ this yields 
\beq\label{NULL}
\frac{{\cal Q}_{0}^{\sigma, \sigma}{\cal Q}_{2}^{\sigma, \sigma}}
{{\cal Q}_{1}^{\sigma, \sigma}{\cal Q}_{1}^{\sigma, \sigma}}=
{\cal Q}_{2}^{\sigma, \sigma}{\cal Q}_{-1}^{\sigma, \sigma}{\cal
  Q}_{-1}^{\sigma, \sigma} = 1
\eeq
which is exactly $\exp - \i \Phi_\sigma$ in the r.h.s. of Eq. (\ref{Q-phases}) for the
translational invariant case. 
For deformations of
the statistics which are not translational invariant, the
spectrum is modified even for a free gas of such particles instead. In the
limit $U\longrightarrow 0$, the phases have to be picked up by proper
convergence of $\sin p_j$ and $\zeta_a$ linearly in $U/t$. 
As a result, $N_\uparrow$ momenta differ from the undeformed values
$p_j^0=\frac{2\pi}{L} k$ by $(\Phi_\uparrow\; {\rm mod}\; 2\pi)/L$ and $N_\downarrow$
momenta differ by $\Phi_\downarrow /L$. 
The energy formula has the form
\begin{equation}
E=-2t\sum_\sigma\sum_{i_\sigma=1}^{N_\sigma} \cos(\frac{2\pi}{L} l_{i_\sigma}+
\frac{\Phi_\sigma}{L}).
\label{desPHYSICAL}
\end{equation}
For finding the ground state,
one has to find the distribution of the distinct integers
$l_{i_\sigma}$ leading to minimal energy. 
This distribution depends on the filling, discriminating less than
quarter filling from fillings in between $1/4$ and $1/2$. Fillings
beyond half filling are to be extracted from the Bethe result
exploiting the particle-hole symmetry of the Hubbard
model~\cite{LIEBWU}.
In the thermodynamic limit, due to the absent symmetry of the momentum
distribution for the ground state, a finite contribution of the
integral over the momenta to order $(1/L)^0$ comes out besides finite
size corrections to order $(1/L)^1$.
Since the energy itself is of order $L$, we will turn to energy
densities, i. e. the energy per site $\eps := \frac{E}{L}$. 
Defining $N_{min}:={\rm min}(N_\uparrow , N_\downarrow )$, 
$\epsilon_\sigma:=\Phi_\sigma \;{\rm mod}\; 2\pi$,
$\epsilon_{cm}:=(\epsilon_\uparrow + \epsilon_\downarrow)/2$, and
${\scriptstyle\Delta}\epsilon:=|\epsilon_\uparrow -
\epsilon_\downarrow|/2$ one obtains for the ground state energy density
$\eps^0_\Phi=:\eps^0_0+\Delta \eps^0$
$$
\Delta \eps^0 = \left\{
\begin{array}{ll}
-\frac{2t}{\pi L}{\scriptstyle\Delta}\epsilon(1-\cos\frac{2\pi N_{min}}{L})+
& \\
+\frac{t}{\pi L^2}(\epsilon_{cm}+{\scriptstyle\Delta}\epsilon)^2
\sin\frac{\pi N}{L} - &; \Phi_\uparrow \Phi_\downarrow < 0 ,
\\
-\frac{2t}{\pi L^2}\epsilon_{cm}{\scriptstyle\Delta}\epsilon
\sin\frac{\pi |N_\uparrow -N_\downarrow |}{L} & \\
&\\
-\frac{2t}{\pi L}{\scriptstyle\Delta}\epsilon (\cos\frac{\pi N}{L}+
\cos\frac{\pi |N_\uparrow -N_\downarrow |}{L})+ & \\
+\frac{t}{\pi L^2}(\epsilon_{cm}+{\scriptstyle\Delta}\epsilon)^2
\sin\frac{\pi N}{L} - &; \Phi_\uparrow \Phi_\downarrow > 0 \\
-\frac{2t}{\pi L^2}(\epsilon_{cm}^2+{\scriptstyle\Delta}\epsilon^2)
\sin\frac{\pi |N_\uparrow -N_\downarrow |}{L} & \frac{1}{4}\leq
\frac{N}{2L}\leq \frac{1}{2} \\ 
&\\
-\frac{2t}{\pi L}{\scriptstyle\Delta}\epsilon (\cos\frac{\pi N}{L}-
\cos\frac{\pi |N_\uparrow -N_\downarrow |}{L})+ & \\
+\frac{t}{\pi L^2}(\epsilon_{cm}+{\scriptstyle\Delta}\epsilon)^2
\sin\frac{\pi N}{L} - &; \Phi_\uparrow \Phi_\downarrow > 0 \\
-\frac{2t}{\pi L^2}(\epsilon_{cm}^2+{\scriptstyle\Delta}\epsilon^2)
\sin\frac{\pi |N_\uparrow -N_\downarrow |}{L} & \frac{N}{2L}< \frac{1}{4}
\end{array}\right .
$$
It is seen that a contribution linear in $1/L$ can occur only for
non-vanishing ${\scriptstyle\Delta}\epsilon$. Otherwise, the effect is
only visible in finite systems to second order in $1/L$.
On the other hand ${\scriptstyle\Delta}\epsilon=0$ also implies that
there is no phase factor in front of the spin part of the Bethe
equation, so that an application of the string hypothesis is
unaffected for this case.

In conclusion, we have shown how DES is  connected to long range correlations 
of  Schulz--Shastry type. The connection  has been revealed realizing operators obeying 
DES through composites of fermionic operators.
From Eqs.~(\ref{desBETHEEQ})~--~(\ref{b-phases}) we conclude that 
solvable DES produces modifications in the boundary 
condition (supporting the general idea of the  
relationship between statistics and topology in $1D$). 
It is important noticing that this is a peculiarity of solvable DES.
DES which are not solvable affect the physics in a different (more complicated) way.
Preliminary results indeed have shown that they modify the spectrum of a 
system with open boundary conditions.
\\   
The characterization
concerning the CBA solvability of the Schulz-Shastry type models\cite{OSAMECK-GEN}
provides the conditions, which the deformation ${\cal Q}$
must fulfill in order that the deformed Hubbard model is CBA
solvable. 
The results obtained in Refs.~\onlinecite{AMOSECK,OSAMECK}
appear as special cases within the class of statistics studied here.
It is worthwhile noting that deformation effects can be seen already
for a free gas of such particles. Its ground state energy as a function
of the deformation parameters is calculated up to second order in
$1/L$, and it contains also a contribution linear in $1/L$ and in the 
asymmetry in the boundary phases $\sDelta\eps$.

We found that correlated deformed exchange statistics 
(with  ${\cal Q}^{\sigma,\sigma'}_{j,k}$ being a
functional of $n_{m,\mu}$)
cannot be realized in terms of fermionic operators. 
Work is in progress along this directions.

\acknowledgements
Motivating discussions with G. Falci, R. Fazio, G. Giaquinta, 
A. Kundu, M. Rasetti, and P. Schwab
are gratefully acknowledged besides the support through the SFB 484
and the {\em Graduiertenkolleg} ``Nonlinear Problems in Analysis,
Geometry, and Physics" (GRK 283), financed by the German Science
Foundation (DFG) and the State of Bavaria.

\end{multicols}
\end{document}